\documentclass[pra,twocolumn,superscriptaddress]{revtex4-2}
\usepackage{graphicx}
\usepackage{amssymb}
\usepackage{amsfonts}
\usepackage{amsmath}
\usepackage{lmodern}
\usepackage{mathrsfs}
\usepackage{mathdots}
\usepackage{collref}
\usepackage[colorlinks=true, citecolor=blue, urlcolor=blue]{hyperref}
\usepackage{tikz}
\usepackage{bm}

\begin{document}

\title{Temperature-heat uncertainty relation in nonequilibrium quantum thermometry}
\author{Ning Zhang}
\affiliation{Lanzhou Center for Theoretical Physics and Key Laboratory of Theoretical Physics of Gansu Province, Lanzhou University, Lanzhou 730000, China}

\author{Si-Yuan Bai}
\affiliation{Lanzhou Center for Theoretical Physics and Key Laboratory of Theoretical Physics of Gansu Province, Lanzhou University, Lanzhou 730000, China}

\author{Chong Chen}\email{chongchenn@gmail.com}
\affiliation{Department of Physics and The Hong Kong Institute of Quantum Information of Science and Technology, The Chinese University of Hong Kong, Shatin, New Territories, Hong Kong, China}

\begin{abstract}
We investigate the temperature uncertainty relation in nonequilibrium probe-based temperature estimation process. We demonstrate that it is the fluctuation of heat that fundamentally determines temperature precision through the temperature-heat uncertainty relation. Specifically, we find that heat is divided into trajectory heat and correlation heat, which are associated with the heat exchange along thermometer's evolution and the correlation between the thermometer and the sample, respectively. Based on two type of thermometers, we show that both of these heat terms are resources for enhancing temperature precision.  By clearly distinguishing the resources for enhancing estimation precision, our findings not only explain why various quantum features are crucial for accurate temperature sensing but also provide valuable insights for designing ultrahigh-sensitive quantum thermometers.  Additionally, we demonstrate that the temperature-heat uncertainty relation is consistent with the well-known temperature-energy uncertainty relation in  thermodynamics. It establishes a connection between the information theory and the thermodynamics.
\end{abstract}
\maketitle
\section{Introduction}
The enhancement of temperature accuracy would help us better quantify and control various physical, chemical, and biological processes that are sensitive to temperature variations \cite{Puglisi2017,Neumann2013,Kucsko2013,FerreiroVila2021}. 
Various quantum features, e.g., quantum coherence \cite{Stace2010, Jevtic2015,Mukherjee2019,Mitchison2020,Zhang2022b,Adam2022,Yuan2023}, strong coupling \cite{Correa2017,Mehboudi2019,Zhang2021,Mihailescu2023,Brenes2023,Xu2023}, quantum correlations \cite{Seah2019, Planella2022}, quantum criticality \cite{Hovhannisyan2018,Mirkhalaf2021,Aybar2022,Zhang2022} are proposed to enhance the temperature measurement precision, which is known as the quantum thermometry \cite{Giazotto2006, Hofer2017, DePasquale2018, Mehboudi2019b, Potts2019, Rubio2021,Mehboudi2022}.  However, the fundamental limits of the quantum thermometry and the reasons why these quantum features can enhance estimation precision are still not fully understood.

Similar to most sensing processes, temperature sensing consists of three steps: the initial state preparation, interact with the sample and evolution, and the final measurement and read out \cite{Escher2011,Giovannetti2011}. Thermometers are divided into two classes based on the evolution process, one is the equilibrium thermometer, where the probe and sample reach thermal equilibrium \cite{Correa2015,DePasquale2016,Hovhannisyan2021}. The other is the non-equilibrium thermometer \cite{Cavina2018, Mukherjee2019, Razavian2019, Mitchison2020,Bouton2020}. The precision for both classes is constrained by the Cram\'{e}r-Rao bound \cite{DePasquale2018}. 
In the equilibrium case, when the coupling between the thermometer and the sample is negligible, this bound is reduced to the well-known temperature-energy uncertainty relation (UR) in thermodynamics, $\Delta \beta \Delta U \ge 1$ \cite{Landau1980, Paris2015}.  It reveals that the precision of temperature is fundamentally relies on the fluctuation of internal energy \cite{Miller2018}. Similar to the position-momentum and time-energy URs, this UR highlights the resources — the fluctuation of internal energy — for enhancing temperature precision \cite{Coles2017}. Based on this UR, the optimal thermometry has been proposed for equilibrium thermometer \cite{Correa2015}. Beyond this regime, although various quantum features, such as strong coupling \cite{Correa2017,Mehboudi2019,Zhang2021,Xu2023,Mihailescu2023,Brenes2023} and quantum correlations \cite{Seah2019, Planella2022}, have been demonstrated to contribute positively to quantum thermometry, no specific UR has been identified.  Therefore, the exact resource for enhancing temperature precision beyond weak coupling equilibrium thermometer is still unknown. 

In this article, we reveal a temperature-heat uncertainty relation (UR) in the temperature sensing process, which fundamentally governs the temperature precision.  This conclusion holds for both equilibrium and non-equilibrium quantum thermometers. A detail analysis shows that the heat can be divided into the trajectory heat and correlation heat, which are associated with the heat exchange along thermometer's two-point quantum trajectory and the correlation between thermometer and sample, respectively. Based on two type of thermometers, we demonstrate that both of these heat terms are important for temperature sensing. Specifically, with the heat-exchange thermometer, we demonstrate that enhancing the trajectory heat through decreasing the detuning or increasing the coupling strength is crucial for the low-temperature thermometer.  On the other hand, we reveal that the dephasing thermometer utilizes the correlation heat as a resource for temperature estimation. Thus, enhancing the correlation between the thermometer and the sample  is important for dephasing thermometer. Our findings not only unambiguously reveal the resources in temperature estimation but also clearly explain why quantum features, such as quantum criticality, strong coupling, correlation, and coherence, are essential resources for enhancing estimation precision.

\begin{figure}[thpb]
  \centering
  \includegraphics[width=0.80\columnwidth]{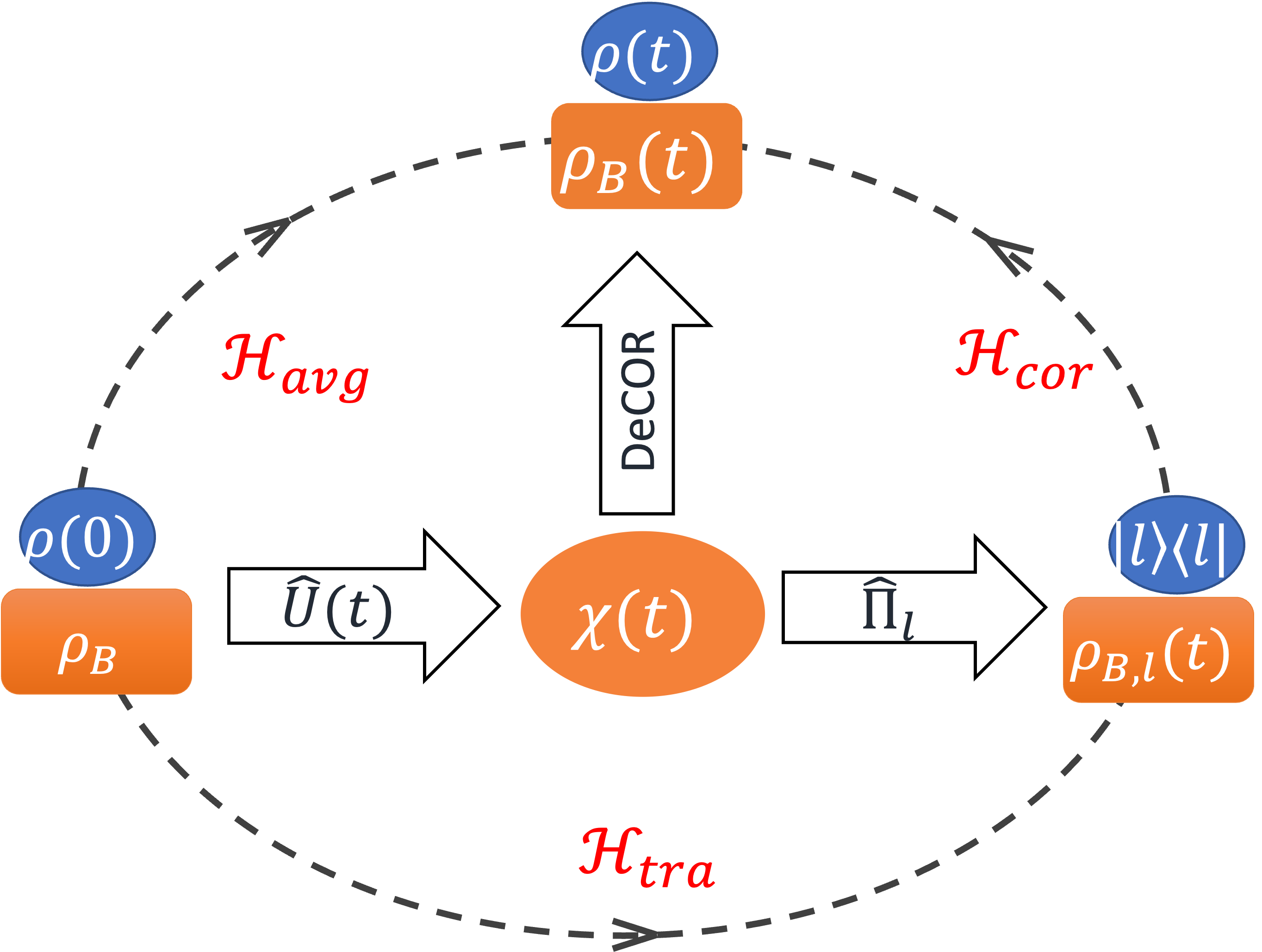}
  \caption{The scheme of temperature estimation. The initial product state $\chi(0)=\rho(0) \otimes \rho_{B}$ evolves to $\chi(t)$ under unitary evolution $\hat{U}_{t}$. A following projection measurement  under a properly chosen basis $\hat{\Pi}_{l}=|l\rangle \langle l|$ is applied on the thermometer part, which reduces the sample state to $\rho_{B,l}(t) = \text{Tr}_S[\hat{\Pi}_l\chi(t)\hat{\Pi}_l]/P_l$ with probability $P_l=\text{Tr}[ \hat{\Pi}_l \chi(t)]$. Sequential measurements on sample's average energy of initial state and final state yields the average energy $\epsilon_{B,l}(0)$ and $\epsilon_{B,l}(t)$, respectively. The difference $\epsilon_{B,l}(0)-\epsilon_{B,l}(t)$ is defined as the trajectory heat as it depends on the two-point quantum trajectory of the thermometer. The sample's average energy is measured from the decorrealated state $\rho_B(t)=\text{Tr}_S[\chi(t)]$, denoted as $\epsilon_{B}(t)$. The disparity between $\epsilon_{B,l}(t)$ and $\epsilon_B(t)$ is defined as the correlation heat, as it reflects the fluctuation of the sample's energy under different final states of the system. At last, the average $\sum_{l} P_l [\epsilon_{B,l}(0)-\epsilon_B(t)]$ over system's all final states is defined as the average heat. In the end, the precision of the temperature estimation is determined by the fluctuation of the term $(\mathcal{H}_{tra}+\mathcal{H}_{cor}-\mathcal{H}_{avg})$.}
  \label{fig:Scheme}
\end{figure}

\section{temperature-heat UR}
Consider a general temperature sensing process. The scheme is illustrated in Fig. \ref{fig:Scheme}. The thermometer (denoted as $S$) is prepared to an initial state $\rho(0)$. The sample (denoted as $B$) is in equilibrium state $\rho_B=e^{-\beta \hat{H}_B}/Z_B$, where $\beta=1/K_B T$ is the inverse temperature with the Boltzmann constant $K_B$ and distribution function $Z_B$. The thermometer then interacts with the sample and evolves according to the total Hamiltonian 
\begin{equation}
	\hat{H}=\hat{H}_{S}+\hat{H}_{B}+\hat{H}_{I},
	\label{eq:Hamil}
\end{equation}
where $\hat{H}_{S}$ is the thermometer's Hamiltonian and $\hat{H}_{I}$ is thermometer-sample  interaction.  After a time $t$ evolution, the total system reads $\chi(t)=\hat{U}_{t}\chi(0) \hat{U}^{\dagger}_{t}$ with $\chi(0)=\rho(0)\otimes \rho_B$ and $\hat{U}_t=e^{-i \hat{H}t}$. Through this evolution process,  the temperature information of the sample is encoded into the thermometer's state $\rho(t)=\text{Tr}_{B}[\chi(t)]$. Properly choosing the measurement basis $\hat{\Pi}_{l}=|l\rangle \langle l|$, the probability distribution  $P_l=\text{Tr}[\hat{\Pi}_{l} \chi(t)]$ is obtained from the projection measurements, which is a function of the temperature. Repeat such processes many times, the temperature is estimated from the probability function $P_l$ and the precision is limited by the Cram\'{e}r-Rao bound
\begin{equation} \label{eq:CRB}
    (\Delta \beta)^2 \ge 1/(N\mathcal{F}), 
\end{equation}
where $N$ is the number of the measurements and $\mathcal{F}$ is the Fisher information with  the definition 
\begin{equation}
    \mathcal{F}=\sum_{l} P_{l} \mathcal{L}^2_{\beta}.
    \label{eq:FI}
\end{equation}
Here $\mathcal{L}_{\beta}=\frac{d \ln P_{l}}{d(-\beta)}$ is the score function \cite{Fisher1925}. Note that different measurement basis $\hat{\Pi}_{l}$ can cause different Fisher information and the optimal basis would yield the largest Fisher information $\mathcal{F}_{Q}$, which is known as the quantum fisher information \cite{Braunstein1994}.

It is interesting to note that the Fisher information can be interpreted as the fluctuation of the score function. Nevertheless, the score function is an information quantity. Here, we aim to find a physical interpretation of the score function such that the information inequality reduces to a physical uncertainty relation. A detailed calculation yields 
\begin{equation}
\mathcal{L}_\beta=\text{Tr}[\hat{M}_l \hat{H}_B \chi(0)\hat{M}^\dagger_{l}]-\text{Tr} [\hat{H}_B \chi(0)],
\label{eq:heat}
\end{equation}
where $\hat{M}_l=\frac{1}{\sqrt{P_l}}\hat{\Pi}_{l} \hat{U}_t$ denotes the free-evolution and the following projection measurement on the system part with output state $| l \rangle$.  From the expression, one can find that the score function is associated with the average energy change of the sample caused by the coupling with the thermometer and the following projection measurement applied on the thermometer.  To correctly capture the physics of the score function, we introduce a decomposition $\mathcal{L}_\beta= \mathcal{H}_{tra} + \mathcal{H}_{cor}-\mathcal{H}_{avg}$, which shows that $\mathcal{L}_\beta$ is the sum of trajectory heat, correlation heat, and average heat \cite{zhang2022c}. The  detail expressions of these heat terms are
\begin{eqnarray}
\label{eq:decomposition-tra}
\mathcal{H}_{tra} &=& \text{Tr}[ \hat{M}_l \hat{H}_B \chi(0)\hat{M}^{\dagger}_l-\hat{H}_B \hat{M}_{l} \chi(0)\hat{M}^{\dagger}_{l}],  ~~\\
\label{eq:decomposition-cor}
\mathcal{H}_{cor} &=&\text{Tr}[\hat{H}_B  \hat{M}_{l}  \chi(0) \hat{M}^{\dagger}_{l}]- \text{Tr}[ \hat{H}_B  \chi(t)],  \\
\label{eq:decomposition-avg}
\mathcal{H}_{avg} &=&\text{Tr}[\hat{H}_B \chi(0)]-\text{Tr}[\hat{H}_B \chi(t)].
\end{eqnarray}
Here, we use the term 'heat' to name the change of sample's average energy \cite{Goold2014,Funo2018,Aurell2018,Popovic2021} although there is still controversy regarding how to define quantum heat in a strong coupling system \cite{Seifert2016,Talkner2016}. 

The physical meaning of these terms is explained via the schematic diagram shown in Fig. \ref{fig:Scheme}. Firstly, the trajectory heat is associated with the average energy change of the sample during the free evolution of the total system, followed by a projection measurement applied to the thermometer.
To check such a point, one can express the Hamiltonian of the sample as $\hat{H}_B=\sum_{m} \varepsilon_{m}|m_B\rangle \langle m_B|$, where $|m_B\rangle$ is the eigenstate of $\hat{H}_B$ with eigenvalue $\varepsilon_m$. Given the initial state $\rho_B=e^{-\beta \hat{H}_B}/Z_B$, one can represent the trajectory heat as the standard heat term in two-point measurements, given by
\begin{equation}
    \mathcal{H}_{tra}=\sum_{m,n} \text{Tr}[\hat{\Pi}^{B}_{n}\hat{M}_{l} \hat{\Pi}^{B}_{m} \chi(0) \hat{\Pi}^{B}_{m} \hat{M}^{\dagger}_{l} \hat{\Pi}^{B}_{n}](\varepsilon_{m}-\varepsilon_{n}),
\end{equation}
where $\hat{\Pi}^{B}_{m}=|m_B\rangle\langle m_B|$ and equalities  $\chi(0)=\sum_{m}\hat{\Pi}^{B}_{m} \chi(0) \hat{\Pi}^{B}_{m}$ and $\hat{H}_B\chi(0)=\sum_{m}\varepsilon_{m}\hat{\Pi}^{B}_{m} \chi(0) \hat{\Pi}^{B}_{m}$ are used.
Twice measurements of $\hat{H}_B$ are conducted on the initial and the final states, yielding the average energy $\epsilon_{B,l}(0)=\sum_{m,n} P_{l;n,m}\varepsilon_{m}$ and $\epsilon_{B,l}(t)=\sum_{m,n} P_{l;n,m}\varepsilon_{n}$, respectively, where  $P_{l;n,m}=\text{Tr}[\hat{\Pi}^{B}_{n}\hat{M}_{l} \hat{\Pi}^{B}_{m} \chi(0) \hat{\Pi}^{B}_{m} \hat{M}^{\dagger}_{l} \hat{\Pi}^{B}_{n}]$ and satisfies $\sum_{m,n}P_{l;n,m}=1$. Their difference is defined as the trajectory heat, which characterizes the loss of sample's average energy along temperature measuring process, tracing the trajectory of the thermometer from  $\rho(0)$ to $|l\rangle \langle l|$. 
Secondly, the correlation heat quantifies the average energy change of the sample caused by the projection measurement applied to the thermometer. The sample's state with and without a projection measurement of the thermometer are represented by $\rho_{B,l}(t)=\text{Tr}_{S}[\hat{M}_{l} \chi(0) \hat{M}^{\dagger}_{l}]$ and $\rho_B(t)=\text{Tr}_S[\chi(t)]$, respectively. Both are density matrices and adhere to the relation $\rho_{B}(t)=\sum_{l} P_l \rho_{B,l}(t)$. The correlation heat is then reduced to 
\begin{equation}
    \mathcal{H}_{cor}=\text{Tr}[\hat{H}_B \rho_{B,l}(t)]-\text{Tr}[\hat{H}_B \rho_{B}(t)].
\end{equation}
The corresponding average energies in above equation are denoted as $\epsilon_{B,l}(t)$ and $\epsilon_{B}(t)$, respectively.  Their difference is defined as the correlation heat $\mathcal{H}_{cor}$. It quantifies the fluctuation of the sample's average energy under different final states $|l\rangle \langle l|$ of the system.  Here, we refer $\mathcal{H}_{cor}$ to correlation heat because it depends on the correlation between thermometer and sample.  In the absence of any correlation between the thermometer and the sample, the measurement applied on the thermometer would not alter the average energy of the sample, resulting in $H_{cor}=0$.  One can check such a point by setting  $\chi(t)=\rho(t) \otimes \rho_{B}(t)$ as a product state, which yields $\rho_{B,l}(t)=\rho_B(t)$. Finally, the difference in sample's average energy between $\sum_{l} P_l \epsilon_{B,l}(0)$ and $\epsilon_{B}(t)$ is defined as the average heat \cite{Popovic2021}.

Noting further that $\mathcal{H}_{avg}=\langle  \mathcal{H}_{tra} \rangle$ is the mean of $\mathcal{H}_{tra}$, we get a simple expression of the score
 \begin{equation}
 \mathcal{L}_\beta = \delta \mathcal{H}_{tra}+\mathcal{H}_{cor},
 \label{eq:L-heat}
  \end{equation}
 where $\delta \mathcal{H}_{tra}=\mathcal{H}_{tra} -\mathcal{H}_{avg}$. Hereafter, we use $\langle \rangle$ to denote the average over $P_l$. Using these results, we find that  $\mathcal{F}= \langle (\delta \mathcal{H}_{tra}+\mathcal{H}_{cor})^2\rangle$. It is the main result of this paper. It reveals that it is the fluctuation of heat that fundamentally determines the temperature precision, i.e.,
\begin{equation}
\Delta \beta^2 \langle (\delta \mathcal{H}_{tra}+\mathcal{H}_{cor})^2 \rangle \ge 1.
\label{eq:GTU}
\end{equation}
Here, we set the number of  measurements $N=1$ for representation convenience. As it shares the same expression as the UR, we refer it as the temperature-heat UR in the temperature sensing process. From it, we can identify two resources to enhance temperature precision: one is the sample's average energy change along the whole temperature measurement process. It characterizes the ability for heat exchange \cite{Potts2019}; the other is sample's average energy change along the projection measurement. It characterizes the correlation between the thermometer and the sample \cite{Seah2019,Planella2022}.  In the following section, we illustrate these two types of resources based on specific thermometers.

\section{heat-exchange thermometer}
The first type of thermometer we considered operates through the exchange of excitations or heat  \cite{Correa2015,Jorgensen2020, Zhang2022}. Here, we take a toy model consisting of two coupled oscillators to illustrate this type of quantum thermometry. One oscillator is employed as the thermometer, while the other functions as the thermal sample. Despite its simplicity, this model captures many key features of thermometers working through heat exchange. The Hamiltonians of $\hat{H}_S$, $\hat{H}_B$, and $\hat{H}_I$ read
\begin{equation}\label{eq:Hamiltonian-EPT}
    \hat{H}_S= \omega_a \hat{a}^{\dagger} \hat{a}, ~~\hat{H}_B=\omega_0 \hat{b}^{\dagger} \hat{b}, ~~ \hat{H}_I=g (\hat{a}^{\dagger} \hat{b}+\hat{b}^{\dagger} \hat{a}),
\end{equation}
where $\hat{a} (\hat{b})$ and $\hat{a}^{\dagger} (\hat{b}^{\dagger})$ are the annihilation and creation operators of the thermometer (sample), respectively, with the characteristic energy $\omega_a (\omega_0)$ and $g$ is the coupling strength. We set the initial state of the thermometer as the vacuum state $\rho(0)=|0_a\rangle\langle 0_a|$, which is proven to be the optimal for temperature estimation \cite{Correa2015}.

Set the Fock state $|l_a\rangle$ as the measurement basis, which is demonstrated to be the optimal basis for temperature estimation \cite{Zhang2022}. One can exactly solve this model and get the trajectory heat and the correlation heat. Detail derivation are shown in Appendix \ref{app:A}. Results are
\begin{equation}\label{eq:Heat-EPT}
    \mathcal{H}_{tra}=l \omega_0 , ~~\mathcal{H}_{cor}= \frac{\bar{n}_b-\bar{n}(t)}{1+\bar{n}(t)} \delta l \omega_0,
\end{equation}
where $\delta l=l-\bar{n}(t)$ with $\bar{n}(t)=\langle l \rangle$ being the mean of $l$ and $\bar{n}_b=\frac{1}{e^{\beta \omega_0}-1}$ is the average excitation number in the initial state of the sample.  Detail calculation yields $\bar{n}(t)=\frac{g^2 \bar{n}_b}{\Delta^2+g^2} \sin^2 \sqrt{\Delta^2+g^2} t$ with the detuning $\Delta=(\omega_a-\omega_0)/2$ and the fluctuation $\langle (\delta l)^2 \rangle=\bar{n}(t)(1+\bar{n}(t))$.  Using these results, we obtain 
\begin{equation}\label{eq:bound-EPT}
    \Delta \beta/\beta \ge \frac{1}{(\beta\omega_0)(1+\bar{n}_b)}\sqrt{\frac{1+\bar{n}(t)}{\bar{n}(t)}}. 
\end{equation}

From Eq. \eqref{eq:Heat-EPT}, it is evident that the trajectory heat is directly proportional to the number of excitations absorbed from the sample, which reflects the ability to exchange heat. To enhance the fluctuation of the trajectory heat, one needs to optimize the heat exchange ability by selecting the optimal evolution time, denoted as $t_{opt}=(i+1/2)\pi/\sqrt{\Delta^2 +g^2}$, where $i\in \mathbb{Z}$ and the optimal detuning, denoted as $\Delta_{opt}=0$. Conversely, a large detuning $\Delta \gg g$ would obstruct the heat exchange, rendering the thermometer inefficient.  Additionally, examining the correlation heat in Eq. \eqref{eq:Heat-EPT}, it is proportional to the excitation number in the sample $\bar{n}_b-\bar{n}(t)$. One can attribute this term to the conservation of the total particle number $\hat{N}=\hat{a}^{\dagger} \hat{a}+\hat{b}^{\dagger} \hat{b}$ due to $[\hat{H},\hat{N}]=0$.  As the correlation heat shares the same sign with the trajectory heat, it  consistently enhances the temperature precision with a factor $(1+\bar{n}(t))/(1+\bar{n}_b)$ .  This enhancement becomes particularly dominant during rapid high-temperature sensing, where $(1+\bar{n}(t))/(1+\bar{n}_b) \ll 1$. Through this simple example, we demonstrate that the ability to exchange heat is the primary resource for heat-exchange quantum thermometers, and the correlation heat further enhances the temperature precision. 

The increase of the heat exchange ability would help us to realize the high-precision quantum thermometer, especially at low-temperature.  In low-temperature quantum thermometry, there is an estimation error divergence problem \cite{Potts2019} caused by the mismatch between thermometer's characteristic frequency  $\omega_a$ and sample's typical frequency $\omega_0\approx K_B T$.  For a multimode sample, Jørgensen et al. revealed a tight bound $(\Delta \beta/\beta) \sim (\beta)^{(1+s)/2}$ on the finite-resolution quantum thermometry, where  the multimode sample is described by the spectral density $J(\omega):=\sum_k g_k^2 \delta(\omega-\omega_k)=\alpha \omega^{s}\omega^{1-s}_c e^{-\omega/\omega_c}$ with $s\ge 0$ \cite{Jorgensen2020}.  One can understand this result from a single mode approximation $g^2 \approx \int^{K_B T}_0 d\omega J(\omega)\sim (K_B T)^{1+s}$ with $\omega_0 \approx K_B T$.  It yields the bound  $\Delta\beta/\beta \ge \frac{(e^{\beta \omega_0}-1)^{3/2}}{\beta\omega_0 e^{\beta\omega_0}} \frac{\Delta}{|g|}\sim \beta^{(1+s)/2}$ according to Eq. \eqref{eq:bound-EPT}, which reproduces the result  in \cite{Jorgensen2020}. To address the error divergence issue, one can either reduce the thermometer's characteristic frequency \cite{Paris2015,Hovhannisyan2018,Zhang2022} or enhance the coupling strength \cite{Correa2017,Mehboudi2019,Jorgensen2020}. This highlights the importance of quantum criticality, which effectively provides a gapless thermometer, and the strong coupling for low-temperature thermometers. 

\section{correlation-heat thermometer}
There is another type of thermometer that operates based on correlation heat, which we refer to as a correlation-heat thermometer. One typical example is the dephasing thermometer \cite{Razavian2019, Mitchison2020,Adam2022,Yuan2023}, which operates based on the dephasing process. Here, we consider a multimode sample coupled with a two-level thermometer. By setting $\hat{H}_S=0$ as it is irrelevant to the dephasing process, the Hamiltonians read 
\begin{equation}\label{eq:}
    \hat{H}_I=\hat{\sigma}_z \sum_k g_k (\hat{b}_k+\hat{b}_k), ~~\hat{H}_B=\sum_k \omega_k \hat{b}^{\dagger}_k \hat{b}_k,
\end{equation}
where $\hat{\sigma}_z$ is the Pauli matrix, $\hat{b}_k$ and $\hat{b}^{\dagger}_k$ are the annihilation and creation operators of the $k-$th bosonic mode in the sample, respectively, with characteristic frequency $\omega_k$, and $g_k$ is the coupling strength. 

We set both the initial state and measurement basis to along the $x$-direction, i.e., $\rho(0)=|+_{x}\rangle\langle +_{x}|$ and $\hat{\Pi}_{\pm}=|\pm_x\rangle \langle \pm_x |$, which are proven to be optimal for temperature estimation \cite{Razavian2019}.  Both the trajectory heat and correlation heat are analytically calculated, leveraging the exact solvability of the system \cite{Breuer2007}. Derivations are shown in Appendix \ref{app:B}. Results are
\begin{equation}\label{eq:Heat-DPT}
    \mathcal{H}_{tra}=Q-\frac{l e^{-\Gamma(t)}}{P_l} Q,~~\mathcal{H}_{cor}=\frac{l e^{-\Gamma (t)}}{P_l} (Q+C),
\end{equation}
where $P_l=\frac{1+l e^{-\Gamma(t)}}{2}$ with $l=\pm$ is the probability of the thermometer being $|l_x \rangle$ state,  $Q=-2\sum_k \frac{g^2_k}{\omega_k} (1-\cos(\omega_k t))$ is the average heat absorbed from the sample, $C=-4\sum_k \frac{g^2_k}{\omega_k} \bar{n}_k(1+\bar{n}_k)(1-\cos(\omega_k t))$, and the decoherence factor $\Gamma(t)=4\sum_k \frac{g^2_k}{\omega^2_k} (2 \bar{n}_k +1) (1-\cos(\omega_k t))$ with $\bar{n}_k=\frac{1}{e^{\beta \omega_k}-1}$. 

Equation \eqref{eq:Heat-DPT} reveals that there is an average heat $\langle \mathcal{H}_{tra} \rangle=Q$ in the dephasing process, although dephasing is seen as a process without heat exchange. Detail analysis reveals that, rather than the heat exchange, it is the interaction term $\hat{H}_I$ that induces the trajectory heat \cite{Popovic2021}. It is noteworthy that this term does not contribute to determining the temperature due to its fluctuation, denoted as $\delta \mathcal{H}_{tra}=-\frac{1}{P_l} le^{-\Gamma(t)} Q$, being cancelled out by the corresponding counter term $\frac{1}{P_l} le^{-\Gamma(t)} Q$ in the correlation heat. Thus, it is the correlation heat $\mathcal{H}^\prime_{cor}=\frac{1}{P_l} l e^{-\Gamma(t)} C$ that finally determines the temperature precision in the dephasing thermometer. A simple calculation shows that the temperature precision satisfies 
\begin{equation}\label{eq:FI-DPT}
    \Delta \beta/\beta \ge \sqrt{e^{2 \Gamma (t)}-1}/(2\beta|C|).
\end{equation}
From it, we can draw the conclusion that the correlation heat is the resource for dephasing thermometers. Enhancing the correlation between the sample and thermometer is essential for achieving precise temperature estimation in this type of thermometers \cite{Seah2019, Planella2022}. By setting the dephasing time as $t^{\ast}_2$, in low-temperature limit, the temperature precision  $(\Delta \beta/\beta) \sim  \beta^{1+s}/\alpha \omega^{1-s}_c (t^{\ast}_2)^2$ is obtained
for the spectral density $J(\omega)=\alpha \omega^s \omega^{1-s}_{c} e^{-\omega/\omega_c}$ \cite{Razavian2019,Yuan2023}. From it, we can find that the coherence time $t^{\ast}_2$ servers as a valuable resource for increasing the correlation heat and thus improving temperature precision \cite{Stace2010, Jevtic2015,Mukherjee2019,Mitchison2020,Zhang2022b,Adam2022,Yuan2023}. 

\section{temperature-energy UR}
Before concluding this manuscript, it is interesting to note that the temperature-heat UR, as shown in Eq. \eqref{eq:GTU}, converges to the temperature-energy UR in the steady state limit. Consider the total system is time-independent, it results in the conservation of the total energy. Therefore, the change in the sample's energy is equivalent to the change in the thermometer's energy plus a modification induced by $\hat{H}_{I}$. According to this conservation law, we have
\begin{equation}\label{eq:traj}
    \mathcal{H}_{tra}=\frac{1}{P_{l}}\text{Tr}[\hat{\Pi}_{l}(\hat{H}^{\prime}_{S}\chi(t)-\hat{U}_t\hat{H}^{\prime}_{S}\chi(0)\hat{U}^{\dagger}_t)],
\end{equation}
where $\hat{H}^\prime_{S}=\hat{H}_{S}+\hat{H}_{I}$. On the other hand, in the steady state limit, we know that the thermometer would relaxes to the equilibrium state $\rho_{s}=\text{Tr}_{B}[\chi_s]$, where $\chi_s=e^{-\beta \hat{H}}/Z$ represents the equilibrium state of the total system. This relaxation occurs under the assumption that the size of the thermometer is much smaller than that of the thermal sample \cite{Jarzynski2004,Miller2018,Talkner2020}. The reduced state $\rho_s$ can be further represented as an effective Gibbs state $\rho_{s}=e^{-\beta \hat{H}^{\ast}_{S}}/Z^{\ast}_{S}$ \cite{Jarzynski2004,Talkner2020}. Here, $\hat{H}^{\ast}_{S}$ is known as the Hamiltonian of mean force, given by
\begin{equation}\label{eq:GibbsS}
    \hat{H}^{\ast}_{S}= -\beta [\ln \text{Tr}_B(e^{-\beta \hat{H}})-\ln \text{Tr}_B(e^{-\beta \hat{H}_B})].
\end{equation}
From it, one can derive thermometer's internal energy as $U_{S}=-\partial_{\beta} \ln Z^{\ast}_S$. Moreover, the internal energy can be represented by the energy operator  $\hat{E}^{\ast}_{S}$, defined from the following equation
\begin{equation}\label{eq:energyOperator}
    \frac{d}{d(-\beta)} e^{-\beta \hat{H}^{\ast}_{S}} =\frac{1}{2}(\hat{E}^{\ast}_{S} e^{-\beta \hat{H}^{\ast}_{S}}+e^{-\beta \hat{H}^{\ast}_{S}} \hat{E}^{\ast}_{S}).
\end{equation}
One can check that $\hat{E}^{\ast}_{S}$ is Hermitian and satisfies $\langle \hat{E}^{\ast}_{S} \rangle=U_{S}$. Importantly, it restores to $\hat{H}_S$ in the weak coupling limit that $\hat{E}^{\ast}_{S}=\hat{H}^{\ast}_{S}=\hat{H}_{S}$. Note that the effective energy operator  $\hat{E}^{\ast}_{S}$ is different from the one $\hat{E}^{\ast}_{S}=\partial_\beta (\beta \hat{H}^{\ast}_S)$ used in Ref. \cite{Miller2018}. However, both two definitions satisfy $\langle \hat{E}^{\ast}_{S} \rangle=U_{S}$ and share the same classical corresponding \cite{Seifert2016}. 

By selecting $\hat{\Pi}_l$ as the eigenstate projector operator of $\hat{E}^{\ast}_{S}$ with eigenvalue $\epsilon_{S,l}$, one can determine the deviation of the internal energy as $\delta U_S=(\epsilon_{S,l}-U_{S})$. The detailed expression is obtained by utilizing Eqs. (\ref{eq:GibbsS},\ref{eq:energyOperator}) (see Appendix \ref{app:C} for detail). Result is
\begin{equation}\label{eq:InteralE-fluc}
\delta U_S=\frac{1}{P_l} \text{Tr}[\hat{\Pi}_l (\hat{H}^{\prime}_S+\hat{H}_B)\chi_{s}]-\text{Tr}[ (\hat{H}^{\prime}_S+\hat{H}_B)\chi_{s}].
\end{equation} 
It is exactly the heat term $(\delta \mathcal{H}_{tra}+\mathcal{H}_{cor})$, noting that $\frac{1}{P_l}\text{Tr}[\hat{\Pi}_{l}\hat{U}_t\hat{H}^{\prime}_{S}\chi(0)\hat{U}^{\dagger}_t]-\text{Tr}[\hat{U}_t\hat{H}^{\prime}_{S}\chi(0)\hat{U}^{\dagger}_t]=0$ and $\chi(t)=\chi_s$ in the steady state limit. Thus, the temperature-heat UR shown in Eq. \eqref{eq:GTU} reduces to the temperature-energy UR 
\begin{equation}
    \Delta \beta \Delta U_S\ge 1,
\end{equation}
where $\Delta U^2_{S}=\langle (\delta U_S)^2 \rangle$.  This result not only demonstrates the consistency of the  temperature-heat UR in temperature estimation process and the well known temperature-energy UR in thermodynamics but also establishes a connection between the information theory and the thermodynamics. 

\section{conclusions}
In summary, we reveal the temperature-heat uncertainty relation (UR) in the temperature estimation process, which puts a fundamental limit for quantum thermometry. We find that the estimation precision is fundamentally determined by the fluctuation of the trajectory heat plus the correlation heat. Based on the two type of thermometers, we demonstrate that both of these heat terms are resources for enhancing temperature precision.  By clearly displaying the resources for enhancing temperature precision, our work is helpful for designing high-precision quantum thermometers. Additionally, we demonstrate that the temperature-heat UR converges to the well known temperature-energy UR in the steady state limit. Thus, this result establishes a connection between estimation theory, or information theory, and thermodynamics, which could be valuable for studying quantum thermodynamics in terms of information theory, especially in the strong coupling regime \cite{Talkner2020}.

The work is supported by the National Natural Science Foundation (Grant No. 12247101), Fundamental Research Funds for the Central Universities (Grant No. 561219028), and China Postdoctoral Science Foundation (Grant No. BX20220138 No. 2022M710063).

\appendix 

\section{Heat-exchange thermometer} \label{app:A}

Here, we aim to derive the quantum trajectory heat and correlation heat for a heat-exchange thermometer shown in main text. The Hamiltonians of the thermometer, sample, and the interaction read
\begin{equation}
\hat{H}_{S}=\omega_a\hat{a}^{\dagger}\hat{a},~~\hat{H}_{B}=\omega_{0}\hat{b}^{\dagger}\hat{b},~~\hat{H}_{I}=g(\hat{a}^{\dagger}\hat{b}+\hat{b}^{\dagger}\hat{a}).\label{eq:Hamiltonian}
\end{equation}
Set the initial state of the thermometer state as the vacuum state $|0_{a}\rangle\langle0_{a}|$ and the initial state of the sample as a thermal equilibrium state $\rho_{B}=e^{-\beta\hat{H}_{B}}/Z_{B}$. The evolution of the total system reads
\begin{equation}
\chi(t)=e^{-i\hat{H}t}\chi(0)e^{i\hat{H}t}.\label{eq:evolution}
\end{equation}
By noting that $\rho_{B}=\sum_{m}\frac{\bar{n}_{b}^{m}}{(1+\bar{n}_{b})^{m+1}}\frac{1}{m!}(\hat{b}^{\dagger})^{m}|0_{b}\rangle\langle0_{b}|(\hat{b})^{m}$,
the evolved state is obtained as 
\begin{align}
\chi(t) & =\sum_{m}\frac{\bar{n}_{b}^{m}}{(1+\bar{n}_{b})^{m+1}}\frac{1}{m!}(\hat{b}^{\dagger}(t))^{m}|0_{a},0_{b}\rangle\langle0_{a},0_{b}|(\hat{b}(t))^{m},\label{eq:evolution-D1}
\end{align}
where $\hat{b}(t):=e^{i\hat{H}t}\hat{b}e^{-i\hat{H}t}$ follows the Heisenberg equation of motion. The solution of $\hat{b}(t)$ is exactly obtained as 
\begin{equation}
\hat{b}(t)=e^{-i\bar{\omega}t}\left[(\cos Et+i\frac{\Delta}{E}\sin Et)\hat{b}-i\frac{g}{E}\sin Et\hat{a}\right],\label{eq:solution-b}
\end{equation}
where $\bar{\omega}=(\omega_a+\omega_{0})/2$ and $E=\sqrt{\Delta^{2}+g^{2}}$ with $\Delta=(\omega_a -\omega_{0})/2$. Applying this result to Eq.(\ref{eq:evolution-D1}), we get 
\begin{align}
\chi(t)  = &\sum_{m}\frac{\bar{n}_{b}^{m}}{(1+\bar{n}_{b})^{m+1}}\frac{1}{m!} \nonumber \\
&(\alpha_{t}^{\ast}\hat{b}^{\dagger}+\beta^{\ast}_{t}\hat{a}^{\dagger})^{m}|0_{a},0_{b}\rangle\langle0_{a},0_{b}|(\alpha_{t}\hat{b}+\beta_{t}\hat{a})^{m},\label{eq:evolution-D2}
\end{align}
where $\alpha_{t}=\cos Et+i\frac{\Delta}{E}\sin Et$ and $\beta_{t}=-i\frac{g}{E}\sin Et$.
Setting the Fock state $|l\rangle$ as the measurement basis, we get the reduced
state $\rho_{B,l}=\langle l|\chi(t)|l\rangle/\text{Tr}_{B}[\langle l|\chi(t)|l\rangle]$ as
\begin{align}
\rho_{B,l}  = &\frac{1}{P_{l}}\sum_{m\ge l}\frac{\bar{n}_{b}^{m}}{(1+\bar{n}_{b})^{m+1}}C_{m}^{l}\left(\frac{|\beta_{t}|^{2}}{|\alpha_{t}|^{2}}\right)^{l}|\alpha_{t}|^{2(m-l)} \nonumber\\ 
& |(m-l)_{b}\rangle\langle(m-l)_{b}|,\label{eq:reducedstate-P}
\end{align}
where $\bar{n}_b=\frac{1}{e^{\beta\omega_0}-1}$ is the average excitation number in the initial state of he sample and the probability is defined as $P_{l}=\text{Tr}_{B}[\langle l|\chi(t)|l\rangle]$. Calculation shows that $P_l=\frac{(\bar{n}(t))^{l}}{(1+\bar{n}(t))^{l+1}}$ with $\bar{n}(t)=\bar{n}_b\frac{g^2}{E^2}|\beta_t|^2$. Using these results, we obtain the trajectory heat and correlation heat. Results are 
\begin{align*}
\mathcal{H}_{tra} & =\frac{\omega_{0}}{P_{l}}\sum_{m\ge l}\frac{\bar{n}_{b}^{m}}{(1+\bar{n}_{b})^{m+1}}l C_{m}^{l}\left(\frac{|\beta_{t}|^{2}}{|\alpha_{t}|^{2}}\right)^{l}|\alpha_{t}|^{2(m-l)} \\
&=l\omega_{0}\\
\mathcal{H}_{cor} & =\frac{\omega_{0}}{P_{l}}\sum_{m\ge l}\frac{\bar{n}_{b}^{m} (m-l)}{(1+\bar{n}_{b})^{m+1}}C_{m}^{l}\left(\frac{|\beta_{t}|^{2}}{|\alpha_{t}|^{2}}\right)^{l}|\alpha_{t}|^{2(m-l)}-\epsilon_{B}  \\
&=\frac{\bar{n}_{b}-\bar{n}(t)}{1+\bar{n}(t)}(l+1)\omega_{0}-\epsilon_{B},
\end{align*}
where $\epsilon_{B}=\frac{\bar{n}_{b}-\bar{n}(t)}{1+\bar{n}(t)}\langle l+1\rangle$
is the average energy of the sample. Thus, the final results of $\delta\mathcal{H}_{tra}$
and $\mathcal{H}_{cor}$ are 
\begin{equation}
\delta\mathcal{H}_{tra}=(l-\bar{n}(t))\omega_{0},~~ \mathcal{H}_{cor}=\frac{\bar{n}_{b}-\bar{n}(t)}{1+\bar{n}(t)}(l-\bar{n}(t))\omega_{0},\label{eq:Heatterms}
\end{equation}
where the equality $\langle l\rangle=\bar{n}(t)$ is used.

\section{dephasing thermometer} \label{app:B}

Consider a dephasing model with a two-level thermometer. The Hamiltonian reads
\begin{equation}
\hat{H}=\sum_{k}g_{k}\hat{\sigma}_{z}(\hat{b}_{k}^{\dagger}+\hat{b}_{k})+\sum_{k}\omega_{k}\hat{b}_{k}^{\dagger}\hat{b}_{k},\label{eq:Hamiltonian-DP}
\end{equation}
where the two-level of the thermometer is described by the Pauli matrix $\hat{\sigma}_{z}$, $\hat{b}_{k}is$ annihilation operator of the $k$-th bosonic bath mode with characteristic energy $\omega_{k}$, and $g_{k}$is the coupling strength. The initial state is set to
\[
\chi(0)=|+_{x}\rangle\langle+_{x}|\otimes\rho_{B},
\]
where $\rho_{B}=e^{-\beta\sum_{k}\omega_{k}\hat{b}_{k}^{\dagger}\hat{b}_{k}}/Z$ is in the equilibrium state with $\beta=\frac{1}{K_{B}T}$ being the inverse temperature. Bath's final states under different projections of system state read
\begin{align*}
\langle+_{z}|\chi(t)|+_{z}\rangle  = &\frac{1}{2}e^{-i(\sum_{k}g_{k}\hat{\sigma}_{z}(\hat{b}_{k}^{\dagger}+\hat{b}_{k})+\sum_{k}\omega_{k}\hat{b}_{k}^{\dagger}\hat{b}_{k})t}\rho_{B} \\ 
&e^{i(\sum_{k}g_{k}\hat{\sigma}_{z}(\hat{b}_{k}^{\dagger}+\hat{b}_{k})+\sum_{k}\omega_{k}\hat{b}_{k}^{\dagger}\hat{b}_{k})t},\\
\langle+_{z}|\chi(t)|-_{z}\rangle  = &\frac{1}{2}e^{-i(\sum_{k}g_{k}\hat{\sigma}_{z}(\hat{b}_{k}^{\dagger}+\hat{b}_{k})+\sum_{k}\omega_{k}\hat{b}_{k}^{\dagger}\hat{b}_{k})t}\rho_{B} \\
&e^{-i(\sum_{k}g_{k}\hat{\sigma}_{z}(\hat{b}_{k}^{\dagger}+\hat{b}_{k})+\sum_{k}\omega_{k}\hat{b}_{k}^{\dagger}\hat{b}_{k})t},\\
\langle-_{z}|\chi(t)|-_{z}\rangle  = & \frac{1}{2}e^{i(\sum_{k}g_{k}\hat{\sigma}_{z}(\hat{b}_{k}^{\dagger}+\hat{b}_{k})+\sum_{k}\omega_{k}\hat{b}_{k}^{\dagger}\hat{b}_{k})t}\rho_{B} \\ 
&e^{i(\sum_{k}g_{k}\hat{\sigma}_{z} (\hat{b}_{k}^{\dagger}+\hat{b}_{k})+\sum_{k}\omega_{k}\hat{b}_{k}^{\dagger}\hat{b}_{k})t}.
\end{align*}
Here, we aim to calculate the trajectory heat $\mathcal{H}_{tra}=\frac{1}{P_{l}}\text{Tr}_{B}[\langle l|\hat{U}_t\hat{H}_{B}\chi(0) \hat{U}^{\dagger}_{t}|l\rangle-\langle l|\hat{H}_{B}\chi(t)|l\rangle]$,
where $\hat{U}_t=e^{-i \hat{H} t}$and $|l\rangle=|\pm_{x}\rangle$ is the measurement basis and $P_{l}=\text{Tr}_{B}[\langle l|\chi(t)|l\rangle]$ is the probability. Before calculate $\mathcal{H}_{tra}$, let us
first calculate these terms 
\begin{align*}
\mathcal{H}_{++} & =\text{Tr}_{B}\langle+|\hat{U}_{t}\hat{H}_{B}\chi(0)\hat{U}^{\dagger}_{t}|+\rangle-\text{Tr}_{B}\langle+|\hat{H}_{B}\chi(t)|+\rangle,\\
\mathcal{H}_{+-} & =\text{Tr}_{B}\langle+|\hat{U}_{t}\hat{H}_{B}\chi(0)\hat{U}^{\dagger}_{t}|-\rangle-\text{Tr}_{B}\langle+|\hat{H}_{B}\chi(t)|-\rangle,\\
\mathcal{H}_{-+} & =\text{Tr}_{B}\langle+|\hat{U}_{t}\hat{H}_{B}\chi(0)\hat{U}^{\dagger}_{t}|-\rangle-\text{Tr}_{B}\langle-|\hat{H}_{B}\chi(t)|+\rangle,\\
\mathcal{H}_{--} & =\text{Tr}_{B}\langle-|\hat{U}_{t}\hat{H}_{B}\chi(0)\hat{U}^{\dagger}_{t}|-\rangle-\text{Tr}_{B}\langle-|\hat{H}_{B}\chi(t)|-\rangle.
\end{align*}
Here, we take the term $\mathcal{H}_{+-}$ as an example to show how to calculate it. Using the detail expression of $\hat{H}$, we can demonstrate that
\[\langle+| \hat{U}^{\dagger}_t\hat{H}_B \hat{U}_t|+\rangle=\sum_k \omega_k (\hat{b}^{\dagger}_k+c^\ast_k(t))(\hat{b}_k+c_k(t)),\]
where $c_k(t)=\frac{g_k}{\omega_k}(1-e^{i\omega_k t})$. Using it, we can derive that 
\begin{align*}
    \mathcal{H}_{+-} &= \text{Tr}_{B}\langle+|\hat{U}_{t} \sum_{k}(-c^\ast_k \hat{b}_k-c_k(t) \hat{b}^\dagger_k)\chi(0)\hat{U}^{\dagger}_{t}|-\rangle \\
    & -\sum_k \omega_k |c_k(t)|^2 P_{+-},\\
                    &= \frac{1}{2}\lim_{\lambda \rightarrow 0} \frac{d}{d\lambda}\text{Tr}[ U_t e^{-\lambda\sum_k\omega_k(c^{\ast}_k \hat{b}_k +c_k \hat{b}_{k})} 
                    \rho_B U^{\dagger}_{-t}]\\
                    &-\sum_k \omega_k |c_k(t)|^2 P_{+-},
\end{align*}
where $U_{t}=\mathcal{T}e^{-i \int^t_{0} d\tau \sum_k g_k (\hat{b}^{\dagger}_{k}e^{i\omega_k t}+h.c.)}$, $\mathcal{T}$ and $\bar{\mathcal{T}}$ indicates the time-ordering and the reverse time-ordering, respectively, and $P_{+-}(t)=\text{Tr}_{B}\langle+|\hat{U}_{t}\chi(0)\hat{U}^{\dagger}_{t}|-\rangle$.  
Such a term can be analytically obtained by notating that $\rho_B$ is a Gaussian state, which yields
\begin{align*}
    \mathcal{H}_{+-} & =  -\frac{1}{2} \exp(-\Gamma(t)) \int_\mathcal{C} d\tau \langle \hat{B}(\tau) \hat{C}\rangle-\sum_k \omega_k |c_k(t)|^2 P_{+-},
\end{align*}
where  $\Gamma(t)=4\sum_{k}\frac{g_{k}^{2}}{\omega_{k}^{2}}\coth\frac{\beta\omega_{k}}{2}(1-\cos\omega_{k}t)$ and $\mathcal{C}$ indicates the contour integral that contains two branches $(0,t)$ and $(t,0)$. Here, we introduce the operators $\hat{B}(\tau)=\sum_k g_k (\hat{b}^{\dagger}_{k}e^{i\omega_k \tau}+h.c.)$ and $\hat{C}=\sum_k \omega_k(c^{\ast}_k(t) \hat{b}_k +c_k(t) \hat{b}^{\dagger}_{k})$. Detail calculation yields
\begin{align*}
    \int_\mathcal{C} d\tau \langle \hat{B}(\tau) \hat{C}\rangle=&-2\sum_k \frac{g^2_k}{\omega_k}|1-e^{i\omega_k t}|^2.
\end{align*}
By notating that $P_{+-}=\frac{1}{2}e^{-\Gamma(t)}$ and $\sum_k \omega_k|c_k(t)|^2=\sum_k \frac{g^2_k}{\omega_k}|1-e^{i\omega_k t}|^2$, we finally get
\begin{equation}
    \mathcal{H}_{+-}= \frac{1}{2}e^{-\Gamma(t)}\sum_k \frac{g^2_k}{\omega_k} |1-e^{i\omega_k t}|^2.
\end{equation}

Using the same method, we get
\begin{align*}
\mathcal{H}_{++}=&\mathcal{H}_{--}=-\sum_{k}\frac{g_{k}^{2}}{\omega_{k}}(1-\cos\omega_{k}t), \\
\mathcal{H}_{+-}= &\mathcal{H}_{-+}^{\ast}=  e^{-\Gamma(t)}\sum_{k}\frac{g_{k}^{2}}{\omega_{k}}(1-\cos\omega_{k}t).
\end{align*}
Choosing the measurement basis along $x$-direction, we derive the trajectory
heat  as
\begin{align}
\mathcal{H}_{tra} & =Q-\frac{l e^{-\Gamma(t)}}{P_{l}}Q.\label{eq:Heat-tra-DP}
\end{align}
where $P_{l}=\frac{1}{2}(1+l e^{-\Gamma(t)})$ is the probability of
the thermometer in $|\pm_{x}\rangle$ with $l=\pm$ and $Q=-2\sum_{k}\frac{g_{k}^{2}}{\omega_{k}}(1-\cos\omega_{k}t)$.
The correlation heat is calculated through the same method. The result is
\begin{align}
\mathcal{H}_{cor} & =\frac{1}{P_{l}}le^{-\Gamma(t)}(Q+C),\label{eq:Heat-cor-DP}
\end{align}
where $C=-4\sum_{k}\frac{g_{k}^{2}}{\omega_{k}}\bar{n}_{k}(1+\bar{n}_{k})(1-\cos\omega_{k}t).$

\section{the deviation of internal energy} \label{app:C}
In this section, we aim to derive the detail expression of derivation of internal energy in the steady state limit. Consider the case that the size of sample is far larger than the size of thermometer. In the steady state limit, the reduced state of the thermometer is given by 
\begin{equation}\label{eq:steadyStat-app}
    \rho_{s}=\text{Tr}_{B}[e^{-\beta \hat{H}}/Z],
\end{equation}
where $\hat{H}$ is the total Hamiltonian and $Z=\text{Tr}[e^{-\beta \hat{H}}]$ is the partition function of the total system. It can be represented as an effective Gibbs state $\rho_{s}=e^{-\beta \hat{H}^{\ast}_{S}}/Z^{\ast}_{S}$ in term of the Hamiltonian of the mean force $\hat{H}^{\ast}_{S}$.  The  definition of $\hat{H}^{\ast}_{S}$ is given by
\begin{equation}\label{eq:MeanForce-app}
    \hat{H}^{\ast}_{S}= -\beta [\ln \text{Tr}_B(e^{-\beta \hat{H}})-\ln \text{Tr}_B(e^{-\beta \hat{H}_B})].
\end{equation}
In the weak coupling limit, we have $\hat{H}^{\ast}_{S}=\hat{H}_{S}$.  Based on this representation, the internal energy of the thermometer is represented as 
\begin{equation} \label{eq:internalE-app}
    U_S=\frac{d}{d-\beta} \ln Z^{\ast}_{S}.
\end{equation}
Generally, the internal energy $U_S$ can be represented as the average of the energy operator $\hat{E}^{\ast}_{S}$, denoted as $U_S=\langle \hat{E}^{\ast}_{S}\rangle$. It is important to note that $\hat{E}^{\ast}_{S}$ is different with the $\hat{H}_S$ in the strong coupling case. Here, we define the energy operator using  the following equation  
\begin{equation}\label{eq:energyO-app}
\frac{d}{d-\beta} e^{-\beta \hat{H}^{\ast}_{S}} =\frac{1}{2}(\hat{E}^{\ast}_{S} e^{-\beta \hat{H}^{\ast}_{S}}+e^{-\beta \hat{H}^{\ast}_{S}} \hat{E}^{\ast}_{S}).
\end{equation}

By applying a trace operation on the both side of the above equation, we can prove the equality $U_S=\langle \hat{E}^{\ast}_{S}\rangle$. This is achieved by further dividing a factor $Z^{\ast}_{S}$ on both sides of above equation.
The concept of energy operator originates from the discussion \textit{"First and Second Law of Thermodynamics at Strong Coupling"} by Udo Seifert, which focuses on a classical system. In this context, it is defined as:
\begin{equation}\label{eq:eq:energyO-class-app}
        E^{\ast}_{S}=\partial_{\beta} (\beta H^{\ast}_{S}).
\end{equation}
This definition is generalized to the quantum case in Ref. [41], where it is given by:
\begin{equation}
    \hat{E}^{\ast}_{S}=\partial_{\beta} (\beta \hat{H}^{\ast}_{S}).
\end{equation}
Here, we use a slightly different definition of the energy operator, as shown in Eq. \eqref{eq:energyO-app}.
However, both definitions satisfy the equality $U_S=\langle \hat{E}^{\ast}_{S}\rangle$ and exhibit the same classical correspondence, as shown in Eq. \eqref{eq:eq:energyO-class-app}.

By selecting the measurement operator $\hat{\Pi}_{l}$ as the eigenstate projector operator of $\hat{E}^{\ast}_{S}$ with eigenvalue $\epsilon_{S,l}$, one can determine the deviation of internal energy as
\begin{equation}\label{eq:IE-fluctuation-app}
    \delta U_S=(\epsilon_{S,l}-U_S).
\end{equation}
From Eq. \eqref{eq:energyO-app}, one can further derive that
\begin{align}
    \delta U_S=& \frac{1}{P_l}\text{Tr}_{S}[\hat{\Pi}_{l} \frac{d}{d-\beta} \frac{e^{-\beta \hat{H}^{\ast}_{S}}}{Z^{\ast}_S}] \nonumber \\
    =& \frac{1}{P_l}\text{Tr}_{S}[\hat{\Pi}_{l} \frac{d}{d-\beta} \text{Tr}_B[\frac{e^{-\beta \hat{H}}}{Z}]] \nonumber\\
    =& \frac{1}{P_l} \text{Tr}[\hat{\Pi}_{l} \frac{d}{d-\beta} \frac{e^{-\beta \hat{H}}}{Z}] \nonumber\\
    =& \frac{1}{P_l} \text{Tr}[\hat{\Pi}_{l}  \hat{H} \chi_s] 
    -\text{Tr}[\hat{H} \chi_s], 
\end{align}
where $P_l=\text{Tr}_{S}[\hat{\Pi}_{l} \frac{e^{-\beta \hat{H}^{\ast}_{S}}}{Z^{\ast}_S}]=\text{Tr}[ \hat{\Pi}_{l} \chi_S]$ and $\chi_s=\frac{e^{-\beta \hat{H}}}{Z}$. 
This is exactly the result as shown in the main text Eq. (\ref{eq:InteralE-fluc}). 

\bibliography{TU}

\end{document}